\magnification=1200
\hsize=16.0truecm
\vsize=23.0truecm
\baselineskip=13pt
\pageno=0

\footline={\ifnum\pageno=0\hfil\else\hss\tenrm\folio\hss\fi}

\def\J{$J/\psi$}
\def\C{c{\bar c}}

\def\lsim{\raise0.3ex\hbox{$<$\kern-0.75em\raise-1.1ex\hbox{$\sim$}}}
\def\gsim{\raise0.3ex\hbox{$>$\kern-0.75em\raise-1.1ex\hbox{$\sim$}}}

\newcount\REFERENCENUMBER\REFERENCENUMBER=0
\def\REF#1{\expandafter\ifx\csname RF#1\endcsname\relax
               \global\advance\REFERENCENUMBER by 1
               \expandafter\xdef\csname RF#1\endcsname
                   {\the\REFERENCENUMBER}\fi}
\def\reftag#1{\expandafter\ifx\csname RF#1\endcsname\relax
               \global\advance\REFERENCENUMBER by 1
               \expandafter\xdef\csname RF#1\endcsname
                      {\the\REFERENCENUMBER}\fi
             \csname RF#1\endcsname\relax}
\def\ref#1{\expandafter\ifx\csname RF#1\endcsname\relax
               \global\advance\REFERENCENUMBER by 1
               \expandafter\xdef\csname RF#1\endcsname
                      {\the\REFERENCENUMBER}\fi
             [\csname RF#1\endcsname]\relax}
\def\refto#1#2{\expandafter\ifx\csname RF#1\endcsname\relax
               \global\advance\REFERENCENUMBER by 1
               \expandafter\xdef\csname RF#1\endcsname
                      {\the\REFERENCENUMBER}\fi
           \expandafter\ifx\csname RF#2\endcsname\relax
               \global\advance\REFERENCENUMBER by 1
               \expandafter\xdef\csname RF#2\endcsname
                      {\the\REFERENCENUMBER}\fi
             [\csname RF#1\endcsname--\csname RF#2\endcsname]\relax}
\def\refs#1#2{\expandafter\ifx\csname RF#1\endcsname\relax
               \global\advance\REFERENCENUMBER by 1
               \expandafter\xdef\csname RF#1\endcsname
                      {\the\REFERENCENUMBER}\fi
           \expandafter\ifx\csname RF#2\endcsname\relax
               \global\advance\REFERENCENUMBER by 1
               \expandafter\xdef\csname RF#2\endcsname
                      {\the\REFERENCENUMBER}\fi
            [\csname RF#1\endcsname,\csname RF#2\endcsname]\relax}
\def\refss#1#2#3{\expandafter\ifx\csname RF#1\endcsname\relax
               \global\advance\REFERENCENUMBER by 1
               \expandafter\xdef\csname RF#1\endcsname
                      {\the\REFERENCENUMBER}\fi
           \expandafter\ifx\csname RF#2\endcsname\relax
               \global\advance\REFERENCENUMBER by 1
               \expandafter\xdef\csname RF#2\endcsname
                      {\the\REFERENCENUMBER}\fi
           \expandafter\ifx\csname RF#3\endcsname\relax
               \global\advance\REFERENCENUMBER by 1
               \expandafter\xdef\csname RF#3\endcsname
                      {\the\REFERENCENUMBER}\fi
[\csname RF#1\endcsname,\csname RF#2\endcsname,\csname
RF#3\endcsname]\relax}
\def\refand#1#2{\expandafter\ifx\csname RF#1\endcsname\relax
               \global\advance\REFERENCENUMBER by 1
               \expandafter\xdef\csname RF#1\endcsname
                      {\the\REFERENCENUMBER}\fi
           \expandafter\ifx\csname RF#2\endcsname\relax
               \global\advance\REFERENCENUMBER by 1
               \expandafter\xdef\csname RF#2\endcsname
                      {\the\REFERENCENUMBER}\fi
            [\csname RF#1\endcsname,\csname RF#2\endcsname]\relax}
\def\Ref#1{\expandafter\ifx\csname RF#1\endcsname\relax
               \global\advance\REFERENCENUMBER by 1
               \expandafter\xdef\csname RF#1\endcsname
                      {\the\REFERENCENUMBER}\fi
             [\csname RF#1\endcsname]\relax}
\def\Refto#1#2{\expandafter\ifx\csname RF#1\endcsname\relax
               \global\advance\REFERENCENUMBER by 1
               \expandafter\xdef\csname RF#1\endcsname
                      {\the\REFERENCENUMBER}\fi
           \expandafter\ifx\csname RF#2\endcsname\relax
               \global\advance\REFERENCENUMBER by 1
               \expandafter\xdef\csname RF#2\endcsname
                      {\the\REFERENCENUMBER}\fi
            [\csname RF#1\endcsname--\csname RF#2\endcsname]\relax}
\def\Refand#1#2{\expandafter\ifx\csname RF#1\endcsname\relax
               \global\advance\REFERENCENUMBER by 1
               \expandafter\xdef\csname RF#1\endcsname
                      {\the\REFERENCENUMBER}\fi
           \expandafter\ifx\csname RF#2\endcsname\relax
               \global\advance\REFERENCENUMBER by 1
               \expandafter\xdef\csname RF#2\endcsname
                      {\the\REFERENCENUMBER}\fi
        [\csname RF#1\endcsname,\csname RF#2\endcsname]\relax}
\def\refadd#1{\expandafter\ifx\csname RF#1\endcsname\relax
               \global\advance\REFERENCENUMBER by 1
               \expandafter\xdef\csname RF#1\endcsname
                      {\the\REFERENCENUMBER}\fi \relax}

%
\def\NP{{ Nucl.\ Phys.\ }}
\def\PL{{ Phys.\ Lett.\ }}
\def\PR{{ Phys.\ Rev.\ }}
\def\PRL{{ Phys.\ Rev.\ Lett.\ }}
\def\ZP{{ Z.\ Phys.\ }}

~ \hfill BI-TP 97/02
\vskip 2.5truecm
\centerline{\bf The Transverse Momentum Dependence of Anomalous
\J~Suppression}
\vskip 1.5truecm
\centerline{D.\ Kharzeev, M.\ Nardi and H.\ Satz}
\bigskip
\centerline{Fakult\"at f\"ur Physik, Universit\"at Bielefeld}
\par
\centerline{D-33501 Bielefeld, Germany}
\vskip 2truecm
\centerline{Abstract:}
\bigskip
In proton-nucleus and nucleus-nucleus collisions up to central $S\!-\!U$
interactions, the $P_T$-dependence of \J~production is determined by
initial state parton scattering and pre-resonance nuclear absorption
(``normal"\J~suppression). The ``anomalous" \J~suppression in $Pb\!-\!Pb$
collisions must reduce the normal $P_T$ broadening, since it occurs
mainly in the central part of the interaction region, where also initial
state parton scattering and nuclear absorption are strongest. We thus
expect for $\langle P_T^2 \rangle$ in $Pb\!-\!Pb$ collisions a
turn-over and decrease with increasing $E_T$.
\vfill
\noindent
BI-TP 97/02\par\noindent
January 1997
\eject
The recently announced ``anomalous" \J~suppression in $Pb\!-\!Pb$
collisions \ref{NA50} has provided new support for the hope that colour
deconfinement can be established at CERN-SPS energy. The estimated
energy densities for such collisions do fall into the deconfinement
regime predicted by lattice QCD \ref{QCD}, so that there is a
theoretical basis for such expectations.
\par
\refadd{Matsui}
\refadd{G-H}
\refadd{KLNS}
\refadd{G-V}
\refadd{Capella}
\refadd{Cassing}
\refadd{KLNS}
\refadd{B-O}
\refadd{Wong}
\refadd{Armesto}
\J~suppression was predicted to signal deconfinement \ref{Matsui}.
However, all \J~production in nuclear collisions up to central $S\!-\!U$
interactions shows only pre-resonance absorption in nuclear matter
(``normal \J~suppression") \refs{G-H}{KLNS} and hence no evidence for
deconfinement. In contrast, $Pb\!-\!Pb$ collisions are found to suffer
an additional (``anomalous") suppression, increasing with centrality
\ref{NA50}. So far, this behaviour cannot be consistently accounted
for in terms of hadronic comovers \ref{KLNS}, in spite of a number of
attempts \refto{G-V}{Cassing}. It is compatible with an onset of
deconfinement \ref{KLNS}\refto{B-O}{Armesto}.
\par
\refadd{Cronin}
\refadd{G-G}
\refadd{Pirner}
\refadd{Blaizot}
\refadd{G-S}
Up to now, the experimental studies of anomalous \J~suppression have
addressed only its centrality dependence. In this note, we want to
consider how such suppression should depend on the \J~transverse
momentum. It is well known that the transverse momenta of secondaries
from hadron-nucleus collisions quite generally show a $p_T$-broadening.
For secondary hadrons, this is the Cronin effect \ref{Cronin}; a
similar behaviour is observed also in Drell-Yan and charmonium
production. The natural basis for all such broadening is initial state
parton scattering, and it was in fact shown some time ago
\refto{G-G}{G-S} that this describes quite well the $p_T$-dependence
observed in \J~production from $p\!-\!A$ to central $S\!-\!U$
collisions.
\par
Consider \J~production in $p-\!A$ collisions, assuming gluon fusion as
the dominant process for the creation of a $\C$ pair. Parametrizing
the intrinsic transverse momentum distribution $f(q_T)$ of a gluon in a
nucleon as
$$
f(q_T)= {1\over \pi \langle q_T^2 \rangle}~ \exp\left\{-{q_T^2\over
\langle q_T^2 \rangle}\right\}, \eqno(1)
$$
we obtain by convolution for the transverse momentum distribution
$F_{pA}(P_T)$ of the resulting \J
$$
F_{pA}(P_T)= {1\over \pi \langle P_T^2 \rangle_{pA} }~
\exp\left\{-{P_T^2\over \langle P_T^2 \rangle_{pA}}\right\}, \eqno(2)
$$
with
$$ \langle P_T^2 \rangle_{pA} = \langle q_T^2 \rangle_A  +
\langle q_T^2 \rangle_p. \eqno(3)
$$
The quantity
$$
\delta_{pA} \equiv \langle P_T^2 \rangle_{pA} - \langle P_T^2
\rangle_{pp} =
\langle q_T^2 \rangle_A - \langle q_T^2 \rangle_p \eqno(4)
$$
is thus a suitable measure for the observed nuclear broadening.
\par
Assume now that in the passage of the projectile proton through the
nuclear target, successive interactions broaden the intrinsic momentum
distribution of the corresponding projectile gluon which will eventually
fuse with a target gluon to form a \J~\refto{G-G}{Blaizot}. If the process
of $P_T$ broadening during the passage is a random walk, then the
relevant parameter of the Gaussian distribution (1) becomes
$$
\delta_{pA} = N_c^A \delta_0, \eqno(5)
$$
where $N_c^A$ is the average number of collisions the projectile
undergoes on its passage through the target up to the fusion point,
and $\delta_0$ the average broadening of the intrinsic gluon
distribution per collision.
\par
In nucleus-nucleus collisions, a corresponding broadening occurs for
both target and projectile gluon distributions; here, however,
measurements at fixed tranverse hadronic energy $E_T$ can determine the
broadening for collisions at a given centrality. Hence at fixed impact
parameter $b$ we have
$$
\delta_{AB}(b) = \langle P_T^2 \rangle_{AB}(b) - \langle
P_T^2\rangle_{pp} = N_c^{AB}(b)~ \delta_0, \eqno(6)
$$
with $N_c^{AB}(b)$ denoting the average number of collisions for
projectile nucleons in the target and vice versa, at fixed $b$.
$N_c^{AB}(b)$ has a maximum at small $b$ and then decreases with increasing
$b$; for a hard sphere nuclear model, it would vanish when $b=R_A+R_B$.
\par
In Glauber theory, the quantity $N_c^{AB}(b)/\sigma$ can be calculated
parameter-free from the established nuclear distributions \ref{deJager};
here $\sigma$ denotes the cross section for the interaction of the
nucleon on its passage through the target. We shall determine $\sigma
\delta_0$ from data, so that $\sigma$ never enters explicitly. Once
$\sigma \delta_0$ is fixed, the broadening by initial state
parton scattering is given for all $p-\!A$ and $A\!-\!B$ interactions.
For Drell-Yan
production (with quarks instead of gluons in the partonic interaction),
this would be the observed effect, since the final state
virtual photon does not undergo any further (strong) interactions.
A produced nascent \J~will, however, experience pre-resonance nuclear
absorption; this suppresses \J's produced early along the path of the
projectile, since they traverse more nuclear matter and hence
are absorbed more than those produced later. As a net result, this
shifts the effective production point to a later stage. In $p-\!A$
collisions, a Drell-Yan pair will on the average be produced in the
center of the target. In contrast,
nuclear absorption shifts the average $\C$ production point further
down-stream. This effectively lengthens the path for initial state parton
scattering and hence increases the resulting broadening.
\par
The transverse momentum behaviour of normal \J~production in nuclear
collisions is thus a combination of initial state parton scattering
before the production of the basic $\C$ state, and pre-resonance
nuclear absorption afterwards; both lead to a broadening of a
$P_T$-distribution. A further broadening could arise from elastic random
walk scattering of the charmonium state itself in nuclear 
matter; however, such an effect will be included here if we fit $\sigma
\delta_0$ to the data.
\par
The essential task is thus to calculate the number of collisions per
cross section, $N_c/\sigma$, for $p-\!A$ and $A\!-\!B$ interactions,
taking into account the effect of pre-resonance nuclear absorption.
\par
We begin with $p-\!A$ collisions. The number of collisions which the
projectile nucleon undergoes up to the $\C$ formation point $(b,z)$
inside the target is given by
$$
{N_c(b,z)\over \sigma} = A~T_A(b,z) = A \int_{-\infty}^z dz' \rho(b,z'),
\eqno(7)
$$
with $b$ denoting the impact parameter. The total effective number of
collisions before $\C$ formation is obtained by averaging over $z$
with the nuclear density $\rho(b,z)$ and the survival probability
$S_A(b,z)$ under nuclear absorption as weights,
$$
{N_c(b)\over \sigma} = \int_{-\infty}^{\infty} dz~ \rho(b,z)~ 
{N_c(b,z)~\over \sigma}~
S_A(b,z) {\bigg /} \int_{-\infty}^{\infty} dz~ \rho(b,z)~
S_A(b,z).
\eqno(8)
$$
The survival probability in Eq.\ (8) is given by
$$
S_A(b,z) = \exp\left\{-(A-1)~\sigma_{abs} \int_z^{\infty} dz'
\rho_A(b,z') \right\}, \eqno(9)
$$
where $\sigma_{abs}$ is the absorption cross section for pre-resonance
charmonium in nuclear matter; in a recent analysis \ref{KLNS}, it was
found to be $7.3 \pm 0.6$ mb. Eqs.\ (8) and (9) lead to
$$
{N_c(b)\over \sigma} = {AT_A(b) \over 1 -
\exp\{-(A-1)\sigma_{abs}T_A(b)\}} - {A \over (A-1) \sigma_{abs}}.
\eqno(10)
$$
From Eq.\ (10) we get
$$
{N_c(b)\over \sigma}{\Big |}_{\sigma_{abs}=0} = {1\over 2} AT_A(b),
\eqno(11)
$$
and
$$
{N_c(b)\over \sigma}{\Big |}_{\sigma_{abs}=\infty} = AT_A(b); \eqno(12)
$$
for no pre-resonance absorption, the formation thus occurs in the center
of the nucleus, for infinite absorption on the far surface, as expected.
\par
Averaging the number of collisions over the impact parameter $b$ then
gives us the required $N_c/\sigma$ for $p-\!A$ interactions,
$$
{N_c\over \sigma} = \int d^2b~ {N_c(b)\over \sigma}~[1-P_0(b)] {\bigg /}
\int d^2b~ [1-P_0(b)]. \eqno(13)
$$
Here $P_0(b)=[1-\sigma T_A(b)]^A$ is the probability for no interaction
of the projectile; for a hard sphere nuclear distribution, it would
become $\Theta(b-R_A)$.
\par
The formulation of the corresponding expressions for $A\!-\!B$
interactions is quite straight-forward. The $\C$ formation point is now
specified by the impact parameter $b$, the positions $(s,z)$ and
$(b-s,z')$ in the two nuclei, with $s$ in the transverse plane and
$z,z'$ along the beam axis. The number of collisions up to the
formation point becomes
$$
{N_c(b,s,z,z')\over \sigma} = A\int_{-\infty}^z dz_A~ \rho_A(s,z_A)
+ B\int_{-\infty}^{z'} dz_B~ \rho_B(b-s,z_B), \eqno(14)
$$
and the corresponding average number of collisions in the presence of
pre-resonance absorption is for fixed impact parameter $b$ given by
\vfill\eject
$$
{N_c(b)\over \sigma} = \int d^2s \int_{-\infty}^{\infty} dz~\rho_A(s,z)
\int_{-\infty}^{\infty} dz'~\rho_B(b-s,z')~ S_A(s,z)~ S_B(b-s,z')~
{N_c(b,s,z,z')\over\sigma}
$$
$$
{\bigg /}\int d^2s \int_{-\infty}^{\infty} dz~\rho_A(s,z)
\int_{-\infty}^{\infty} dz'~\rho_B(b-s,z')~ S_A(s,z)~ S_B(b-s,z').
\eqno(15)
$$
From this, we can in turn obtain the corresponding value at fixed
transverse energy $E_T$ in the usual fashion,
$$
{N_c(E_T) \over \sigma} = \int d^2b~P(E_T,b)~[1-P_0(b)]~{N_c(b)\over
\sigma} {\bigg /}
\int d^2b~P(E_T,b)~[1-P_0(b)], \eqno(16)
$$
by convolution with the $E_T-b$ correlation function $P(E_T,b)$
\ref{KLNS}. $P_0(b)$ here denotes the probability for no
interaction in $A\!-\!B$ collisions, a generalisation of the $p-\!A$
form used above.
\par
With Eq.\ (13) for $p-\!A$ and Eqs.\ (15/16) for $A\!-\!B$ collisions,
we have the required Glauber results. Ideally, we would now use
Eqs.\ (4/5) and $p-\!A$ data to fix $\sigma \delta_0$; the broadening
for $A\!-\!B$ interactions would then be fully predicted. Unfortunately
there are $p-\!A$ data only for three values of $A$ \refs{NA3}{NA38}, and
these have rather large errors. We shall therefore instead check if
we can obtain a consistent description of all existing $p-\!A$
\refs{NA3}{NA38} and $A\!-\!B$ data \refs{NA38}{Mandry}, up to central
$S\!-\!U$, in terms of a common
$\sigma \delta_0$. For $\langle P_T^2 \rangle_{pp}$, NA3 data on
\J~production at 200 GeV beam momentum \ref{NA3} give $1.23\pm 0.05$
GeV$^2$; this value is confirmed in an NA38 analysis \ref{NA38}
as well as by a two-parameter fit to all data which we have performed.
Using the NA3 proton-proton value, we obtain the best fit to the
available
$p-\!A$, $O\!-\!Cu$, $O\!-\!U$ and $S\!-\!U$ data with $\sigma\delta_0
= 9.4 \pm 0.7$; the error corresponds to 95\% c.\ l., and the minimum
$\chi^2/d.f.$ is 1.1. In the Table, we show the 
experimental results together
with the broadening as obtained from our calculations, using the
mentioned $\sigma_{abs}=7.3\pm0.6$ mb for the pre-resonance absorption
cross section. The behaviour of $\langle P_T^2 \rangle_{SU}$ as
function of $E_T$ is shown in Fig.\ 1. It is seen that initial state
parton scattering and pre-resonance absorption indeed account quite well
for the observed $E_T$ dependence.
\par
We now turn to $Pb\!-\!Pb$ collisions; the corresponding ``normal"
transverse momentum behaviour is shown in Fig.\ 2.\footnote{*}{A similar
form was recently obtained from initial state parton scattering only,
neglecting pre-resonance absorption \ref{G-V/t}. Since such
absorption can be simulated by choosing a somewhat larger $\delta_0$,
the overall pattern remains comparable.} Its basic feature remains the
monotonic increase of $\langle P_T^2 \rangle$ with $E_T$, even though
the collision geometry makes this slightly weaker for $Pb\!-\!Pb$ than
for $S\!-\!U$ interactions \ref{G-V/t}.
\par
The onset of anomalous suppression results in a striking modification of
this pattern. If the \J's in the hot interior of the medium produced in
the collision are suppressed, then this will reduce their contribution
from the part of phase space leading to the most broadening. To
illustrate the effect, we assume suppression by deconfinement; in this
case the result is readily calculable \refss{KLNS}{B-O}{G-S}. To be
specific, we assume deconfinement to start once the interaction measure
\ref{KLNS}
$$
\kappa(b,s) \equiv {N_c(b,s) \over N_w(b,s)} \eqno(17)
$$
passes a certain critical value $\kappa_c$; $N_c$ and $N_w$ are the
number of collisions and of wounded nucleons, respectively, as obtained
from Glauber theory. To calculate the effect of such a melting in the
hot center, the integrations in Eq.\ (15) just have to be constrained
to the cool outer region by introduction of $\Theta(\kappa_c -
\kappa(b,s))$. In Fig.\ 2 we have included the resulting patterns for
several value of $\kappa_c$; the change of behaviour due to anomalous
\J~suppression is qualitatively evident.
\par
To obtain a quantitative prediction for the forthcoming $Pb\!-\!Pb$
results, we take into account the mixed (40\%/60\%) origin of the
observed \J's from intermediate $\chi_c$'s and direct $\psi$
production. In this case, the overall anomalous suppression had been
found \ref{KLNS} to be quite well reproduced with $\kappa_c^{\chi}
\simeq 2.3$ and $\kappa_c^{\psi} \simeq 2.9$. Using these values and the
mentioned $\chi/\psi$ composition, we compare in Fig.\ 3 the anomalous
$E_T$-dependence of $\langle P_T^2 \rangle$ to the normal form.
Deconfinement should thus lead to an onset of anomalous behaviour in the
$P_T$-dependence of \J~production just as it occurs in the integrated
production rates. -- In contrast, the $P_T$ broadening observed for
Drell-Yan production in $p-\!A$ and $A\!-\!B$ collisions
\refto{NA3}{Mandry} should continue in its ``normal" fashion also
for $Pb\!-\!Pb$ interactions.
\par
The observation of such anomalous transverse momentum behaviour
for \J~suppression would thus constitute another piece of evidence
for the onset of new physics in $Pb-Pb$ collisions. In particular, its
occcurrence in $Pb\!-\!Pb$ collisions, with a normal $P_T$-dependence in
$S\!-\!U$ interactions, would again rule out a hadronic explanation of
anomalous \J~suppression: any absorption by hadronic comovers must be
present in continuously varying degrees in all nucleus-nucleus
collisions. On the other hand, as we have seen here, such anomalous
$P_T$-behaviour would arise naturally in case of deconfinement.
It should disappear only for $P_T$ values so high that gluon
fragmentation replaces gluon fusion as dominant \J~production mechanism
or when the relative \J-gluon momentum becomes too large for an effective
\J~dissociation (for $P_T \geq 5 - 10$ GeV) \ref{Wang}.
\vfill\eject
\centerline{\bf Table Caption}
\medskip\noindent
Experimental and theoretical values of $\langle P_T^2 \rangle$ for $p-\!A$ 
and $A\!-\!B$ interactions. Data for $p-p$ and $p-\!Pt$: NA3 \ref{NA3};
the other from NA38 \refs{NA38}{Mandry}.
\bigskip
\centerline{\bf Figure Captions}
\medskip
\noindent
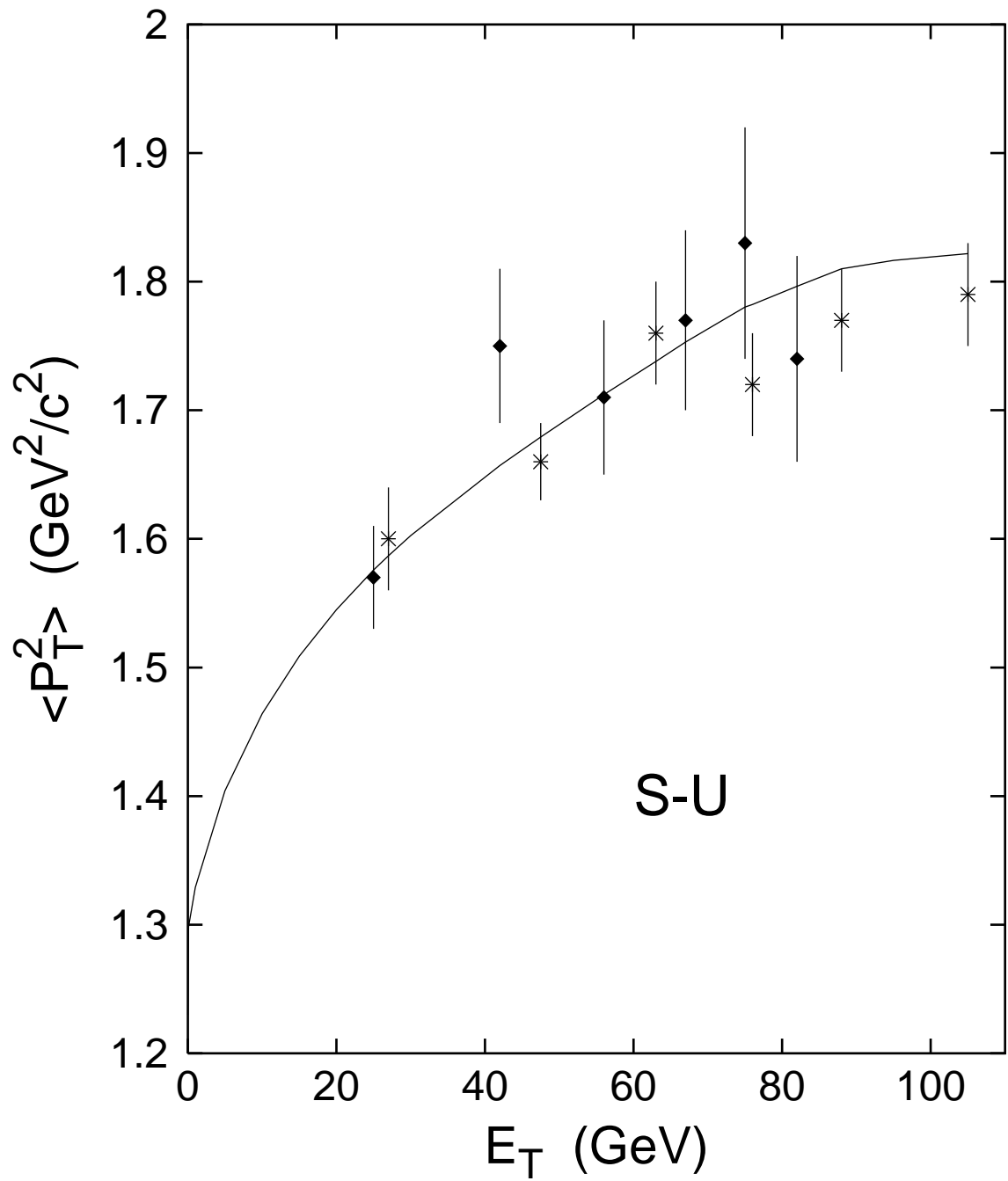
Fig.\ 1: $P_T$ broadening in $S\!-\!U$ collisions; diamonds from \ref{NA38},
stars from \ref{Mandry}.
\medskip
\noindent
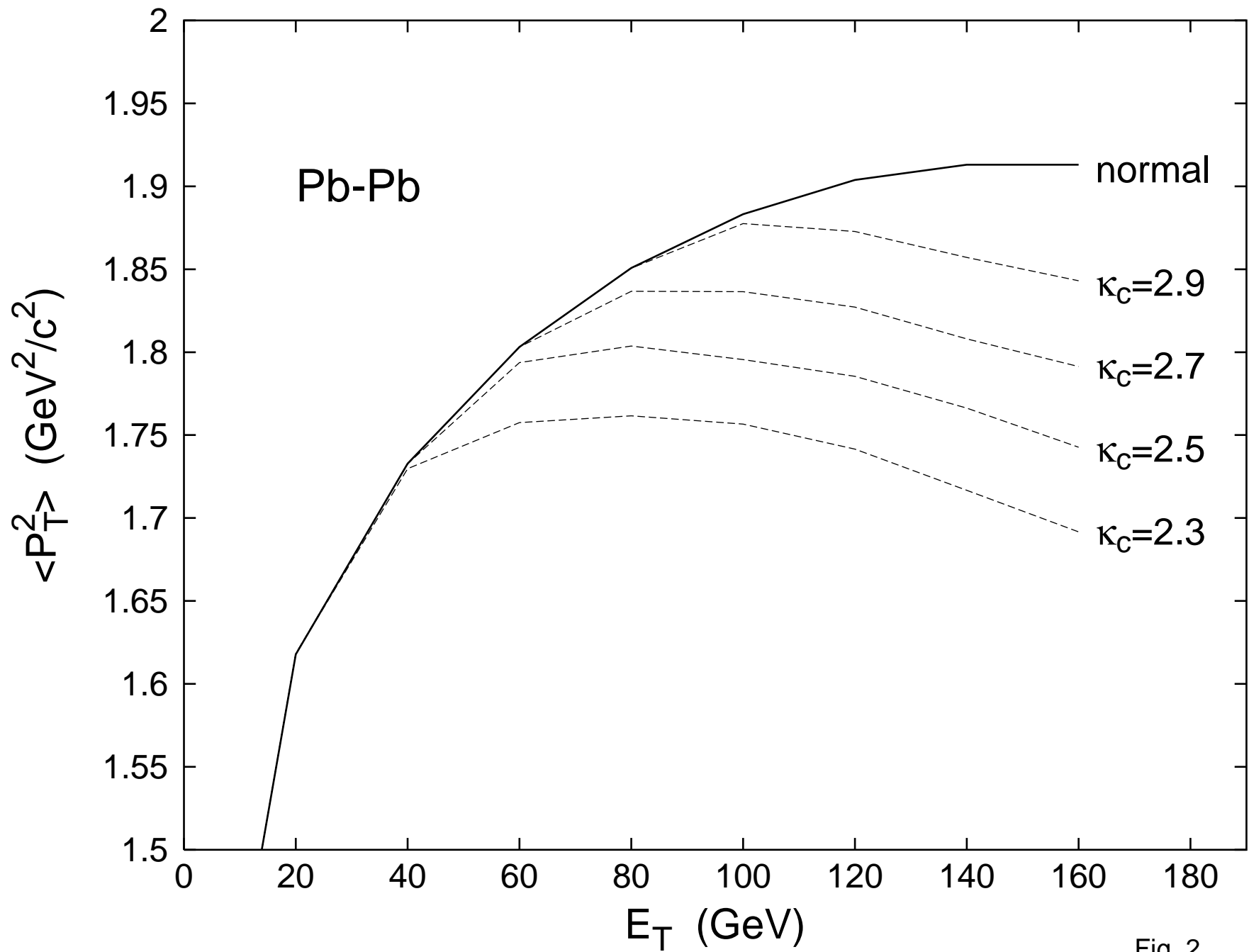
Fig.\ 2: Normal and anomalous $P_T$ behaviour for $Pb\!-\!Pb$ collisions;
the anomalous behaviour is calculated for deconfinement at the given 
critical values of the interaction measure $\kappa_c$.
\medskip
\noindent
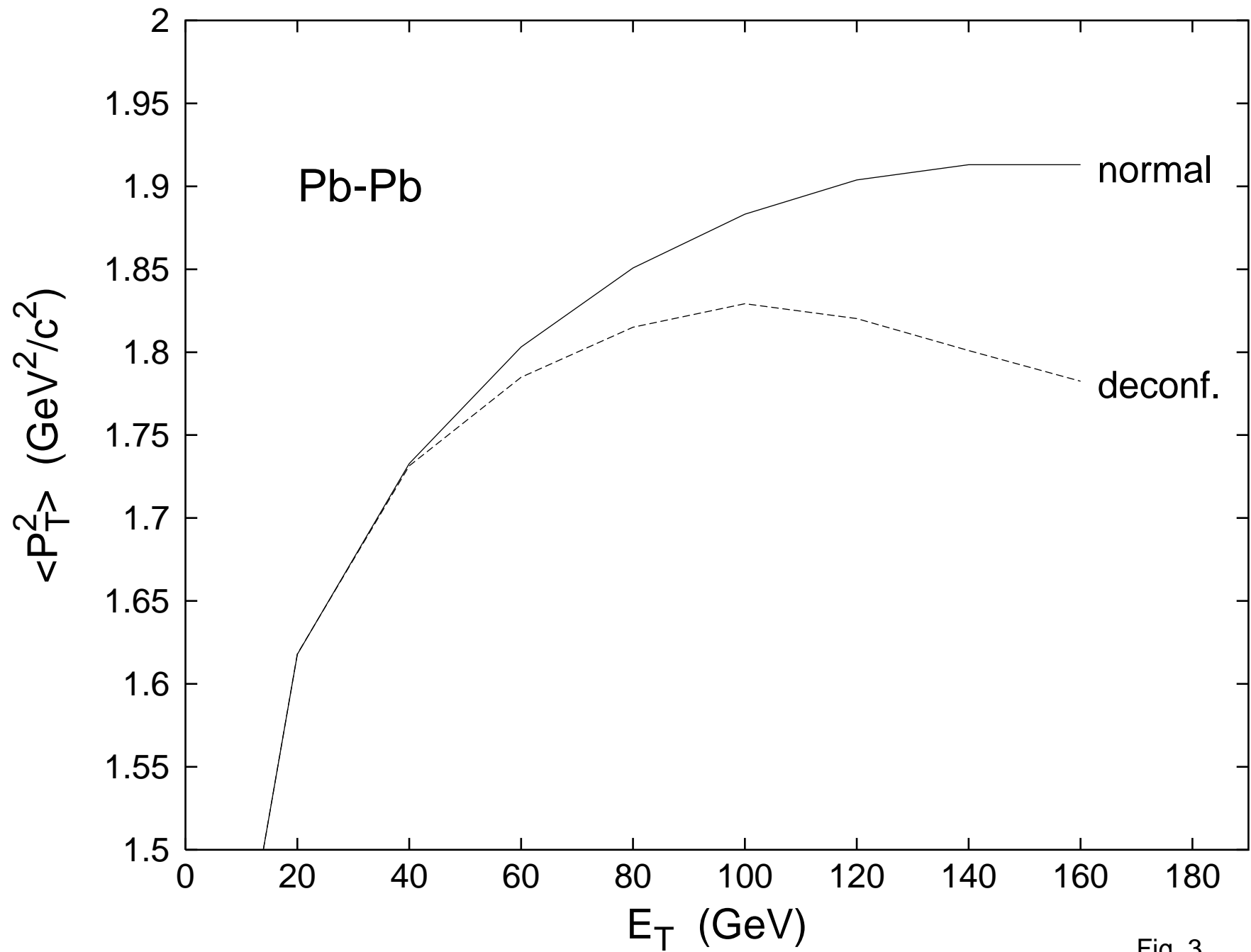
Fig.\ 3: The $P_T$ behaviour for deconfinement in $Pb\!-\!Pb$ collisions
for a 40\%/60\% $\chi/\psi$ origin of the observed \J's, compared to 
normal behaviour.
\bigskip\bigskip

\centerline{\bf References}
\bigskip
\item{\reftag{NA50})}{M.\ Gonin, Report at {\sl Quark Matter 1996},
Heidelberg, Germany;\par
C. Louren\c co,
Report at {\sl Quark Matter 1996}, Heidelberg, Germany.}
\par
\item{\reftag{QCD})}{See e.g., F.\ Karsch, \NP A590 (1995) 367c;\par
F.\ Karsch and H.\ Satz, \ZP C 51 (1991) 209.}
\par
\item{\reftag{Matsui})}{T.\ Matsui and H.\ Satz, \PL 178B (1986) 416.}
\par
\item{\reftag{G-H})}{C.\ Gerschel and J.\ H\"ufner, \ZP C 56 (1992) 171.}
\par
\item{\reftag{KLNS})}{D. Kharzeev et al.\ hep-ph/9612217, Nov. 1996;
\ZP C, in press.}
\par
\item{\reftag{G-V})}
{S.\ Gavin, Report at {\sl Quark Matter 1996}, Heidelberg, Germany;\par
S. Gavin and R. Vogt, hep-ph/9606460, June 1996.}
\par
\item{\reftag{Capella})}{A.\ Capella et al., hep-ph/9607265, July 1996.}
\par
\item{\reftag{Cassing})}
{W.\ Cassing and C.-M.\ Ko, nucl-th/9609025, Sept.\ 1996.}
\par
\item{\reftag{B-O})}
{J.-P.\ Blaizot, Report at {\sl Quark Matter 1996}, Heidelberg, Germany;
\par J.-P. Blaizot and J.-Y. Ollitrault, \PRL 77 (1996) 1703.}
\par
\item{\reftag{Wong})}{C.-Y.\ Wong, hep-ph/9607285, JUly 1996.}
\par
\item{\reftag{Armesto})}
{N. Armesto et al., hep-ph/9607239, July 1996}
\par
\item{\reftag{Cronin})}{J.\ W. Cronin et al., \PR D11 (1975) 3105.}
\par
\item{\reftag{G-G})}{S.\ Gavin and M.\ Gyulassy, \PL 214B (1988) 241.}
\par
\item{\reftag{Pirner})}{J.\ H\"ufner, Y.\ Kurihara and H.\ J.\ Pirner,
\PL 215B (1988) 218.}
\par
\item{\reftag{Blaizot})}{J.-P.\ Blaizot and J.-Y.\ Ollitrault,
\PL 217B (1989) 386 and 392.}
\par
\item{\reftag{G-S})}{S.\ Gupta and H.\ Satz, \PL B 283 (1992) 439.}
\par
\item{\reftag{deJager})}{C.\ W.\ deJager et al.,
Atomic Data and Nuclear Data Tables 14 (1974) 485.}
\par
\item{\reftag{NA3})}{J.\ Badier et al.\ (NA3), \ZP C 20 (1983) 101.}
\par
\item{\reftag{NA38})}{C. Baglin et al.\ (NA38), \PL B 262 (1991) 362.}
\par
\item{\reftag{Mandry})}{R.\ Mandry, Doctorate Thesis, Universit\'e
Claude Bernard, Lyon, November 1993.}
\par
\item{\reftag{G-V/t})}{S.\ Gavin and R.\ Vogt, hep--ph/9610432, Oct.
1996.}
\par
\item{\reftag{Wang})}{X.-M. Xu et al., \PR C53 (1996) 3051.}
\vfill\bye